\documentclass[]{aa}  

\usepackage{graphicx}
\usepackage{natbib}
\usepackage[varg]{txfonts}
\begin{document}
   \title{Star-planet magnetic interaction and evaporation of planetary atmospheres}

   \subtitle{}

   \author{A.~F.~Lanza}

   \institute{INAF-Osservatorio Astrofisico di Catania, Via S.~Sofia, 78 -- 95123 Catania, Italy\\
              \email{nuccio.lanza@oact.inaf.it}
             }

   \date{Received ...; accepted ...}

 
  \abstract
   {Stars interact with their close-in planets  through radiation,  gravitation and magnetic fields.}
   {We investigate the energy input to a planetary atmosphere  by  reconnection between  stellar and  planetary magnetic fields  and compare it to the energy input of the extreme ultraviolet (EUV) radiation field of the star. }
   {We quantify the power released by magnetic reconnection at the boundary of the planetary magnetosphere that is conveyed to the atmosphere by accelerated electrons. We introduce simple models to evaluate the energy spectrum of the accelerated electrons and the energy dissipated in the atmospheric layers in the polar region of the planet upon which they impinge. A simple transonic isothermal wind flow along  field lines is considered to estimate the increase in  mass loss rate in comparison with   a planet irradiated only by the EUV flux of its host star.}
   {We find that energetic electrons can reach levels down to  column densities of $10^{23}-10^{25}$~m$^{-2}$, comparable with or deeper than EUV photons, and increase the mass loss rate up to a factor of $30-50$ in close-in ($ < 0.10 $~AU), massive ($ \geq 1.5 $ Jupiter masses) planets. Mass loss rates up to $(0.5-1.0) \times 10^{9}$~kg~s$^{-1}$ are found for atmospheres heated by electrons accelerated by magnetic reconnection at the boundary of planetary magnetospheres. On the other hand, average mass loss rates up to $(0.3-1.0) \times 10^{10}$~kg~s$^{-1}$ are found in the case of magnetic loops interconnecting the planet with the star.}
   {The star-planet magnetic interaction provides a remarkable source of energy for planetary atmospheres, generally comparable with or exceeding that of  stellar EUV radiation  for close-in planets. Therefore, it must be included in models of chemical evolution or  evaporation of planetary atmospheres as well as in modelling of light curves of transiting planets at UV wavelengths. }


   \maketitle
%

\section{Introduction}
\label{intro}

Several hundreds of extrasolar planets are presently known, about 20 percent of which are Jupiter-mass objects orbiting their host stars within 0.1~AU (hot Jupiters, hereafter HJs). Since late-type stars have outer convection zones that host hydromagnetic dynamos, their  magnetic fields are expected to interact with their HJs, especially if the magnetic field of the planets is of the order of a few Gauss or greater. \citet{Shkolniketal05,Shkolniketal08} found evidence of a chromospheric hot spot rotating in phase with the orbital period of the HJ in  \object{HD~179949} and \object{$\upsilon$~And}  during some observing seasons, while in other seasons
the chromospheric flux showed only the usual rotational modulation. For instance, observing $\upsilon$~And in 2009, \citet{Poppenhaegeretal11} found only a rotational modulation of the chromospheric flux. Absence of a  planetary signature in the case of HD~179949 is reported by \citet{Scandariatoetal12} in the fall of 2009, while \citet{Gurdemiretal12} claim the presence of a planetary-induced hot spot in 2006. Therefore, the chromospheric flux enhancement associated with the planet is a non-stationary phenomenon, observed only in $\approx 30-40$ percent of the seasons. The power irradiated by the hot spot falls in the range $10^{20}-10^{21}$~W \citep{Shkolniketal05}, i.e.,  a few percent of the total chromospheric flux. 

Evidence of a corresponding enhancement in the corona has been provided by \citet{Saaretal08} who observed a modulation of the X-ray emission with the orbital phase in HD~179949. \citet{Kashyapetal08} claimed that stars with massive planets closer than $0.15$~AU have an X-ray coronal luminosity $\approx 2$ times greater than stars with distant planets ($> 1.5$~AU). 
However, \citet{Poppenhaegeretal10} and \citet{PoppenhaegerSchmitt11} questioned such a result and explained the apparent correlation between HJs and enhanced X-ray emission as a consequence of selection effects. They investigated the X-ray spectral energy distribution of six planet-hosting stars  finding that most of the emitting plasma was at temperatures of $\approx 1$~MK, although two stars showed also a significant contribution from a hotter component at $\approx 4-5$~MK. 
\citet{Shkolnik13} has investigated samples of stars with close-in ($a <0.1$~AU) and far-out ($0.5 < a < 2$~AU) planets finding marginal evidence (i.e., at a $2\sigma$ level) of an enhanced far ultraviolet emission in the former. 

Evidence of photospheric signatures associated with a possible magnetic star-planet interaction  are reviewed by, e.g., \citet{Lanza11} considering also their possible connection with  chromospheric and coronal  signatures. 
Recently, \citet{Herreroetal13} have discussed further observations supporting the phenomenon. 

Theoretical models to account for  chromospheric hot spots have been proposed by, e.g., \citet{McIvoretal06}, \citet{Preusseetal06}, and \citet{Lanza08}. Those models do not address the  origin of the energy dissipated in the hot spots limiting themselves to explaining the phase lag between the spots and the planets. The numerical simulations by \citet{Cohenetal11a,Cohenetal11b} indicate that a sufficient power can be released in the case of active stars, such as HD~189733, while \citet{Lanza09,Lanza12} suggests that energy can be released not only at the site where coronal and planetary fields reconnect, but also in the whole interconnecting loops that can  be formed after reconnection (see Sect.~\ref{star-planet_inter} below). 

The outer atmospheres of HJs can be probed through observations of  transiting planets. Specifically, the depth of transit varies as a function of wavelength when an extended atmosphere absorbs the stellar flux. For HD~209458 and HD~189733, the absorption in the Lyman-$\alpha$ line indicates an atmosphere extending beyond the Roche lobe and evaporating from the planet \citep[e.g., ][]{Vidal-Madjaretal03,Lecavelierdesetangsetal10}. 
Recent observations by \citet{Lecavelierdesetangsetal12} reveal a strong temporal variability of the evaporation rate in the case of the active star HD~189733, probably produced by a variation of the extreme ultraviolet (EUV) flux of the star. Atoms and ions of several chemical species, e.g., O, C, Si, Na, have also been observed  \citep[e.g., ][]{Singetal08,Linskyetal10,Vidal-Madjaretal11}. The estimated mass loss rate ranges between $10^{6}$ and $10^{8}$~kg~s$^{-1}$ and the velocity of the evaporation flow is of the order of $\sim 10$~km~s$^{-1}$ with a temperature of $\approx (1-2) \times 10^{4}$~K close to the planet, although velocities up to $\sim 100-150$~km~s$^{-1}$ and temperature of $\approx 10^{5}$~K have been inferred at several planetary radii  \citep[e.g., ][]{Lecavelierdesetangsetal08,Linskyetal10,Lecavelierdesetangsetal12}. With these evaporation rates, a planet having a mass comparable with that of Jupiter can survive till the end of the main-sequence evolution of its host star. However, the situation can be different for less massive planets that experience an higher heating rate during the early main-sequence evolution of their host stars, especially in the first $0.5-1$~Gyr when they are remarkably more active. \citet{Lecavelierdesetangs07} has estimated the lifetime of planets of different mass finding that the observed planets are massive enough to survive evaporation, while no planet is observed in the range of mass that would imply a complete evaporation within a time interval of $\sim 5$~Gyr.

The situation can change if the planet has a magnetic field comparable with that of Jupiter or larger because the field can  trap the ionized gas that is collisionally coupled to the neutral atoms thus preventing the evaporation of the atmosphere. \citet{Trammelletal11} and \citet{Adams11} have investigated the effect of the planetary field on the evaporation considering a simple isothermal wind model. Assuming that the planet has a dipole-like field with the axis of symmetry coincident with the spin axis and normal to the orbital plane, they find that the field prevents evaporation in the low-latitude region where the hot plasma is frozen to closed field lines. At high latitudes, there are open field lines, but the effect of  the tidal force can still reduce or even stop the evaporation in extreme cases \citep{Trammelletal11}. 
Neglecting the effects of  tidal forces, the plasma can freely evaporate at high latitudes, i.e., for a colatitude $\theta \leq \theta_{0}$, where $\theta_{0} \simeq 20^{\circ}-30^{\circ}$  in the model of \citet{Adams11}. In those models, the energy to accelerate the planetary wind  comes from the EUV radiation emitted by the host star. However, the flux reaching the atmospheric layers from which the wind is launched is significantly reduced at high latitudes owing to the oblique trajectory of the light rays. The dilution is proportional to $\sin \theta$ in comparison with  normal incidence. Therefore, it is questionable whether a wind coming from the polar caps of the planet and powered by the stellar EUV flux  can account for the observed mass loss rates \citep[see Sect.~5 in ][]{Adams11}. 

In the present work, we investigate an additional source of energy to power the evaporation flow, that is  non-thermal electrons accelerated at the reconnection site between the stellar and  planetary fields. By analogy with solar flares, non-thermal  electrons  convey  a significant fraction of the energy released in the reconnection process  to the footpoints of the  field lines  in the polar regions of the planet. We consider the power released by magnetic reconnection and compare it with the EUV stellar flux in Sect.~\ref{estimateofevapo}. In Sect.~\ref{model}, we introduce a simple model for the  electron acceleration,  consider the dissipation of energy in the outer planetary atmosphere, and estimate the evaporation rate. In Sect.~\ref{results}, we present an application to 
the sample of transiting planets and discuss the results in Sect.~\ref{discussion}. 

\section{Importance of magnetic-powered evaporation}
\label{estimateofevapo}

The minimum power needed to support an evaporation rate of $\dot{M}_{\rm p}$  can be estimated from the work done  against the planet gravity  as $G M_{\rm p} \dot{M}_{\rm p}/R_{\rm p}$, where $G$ is the gravitation constant, $M_{\rm p}$ the mass of the planet, and $R_{\rm p}$ its radius. If an EUV flux $F_{\rm EUV}$ induces  the evaporation, the  available power is:
\begin{equation}
P_{\rm EUV} = \pi R_{\rm p}^{2} F_{\rm EUV}, 
\label{p-euv}
\end{equation}
that allows us to estimate the maximum evaporation rate as  \citep[cf., e.g., ][]{Watsonetal81,Murray-Clayetal09}:
\begin{equation}
\dot{M_{\rm p}} = \frac{\pi R_{\rm p}^{3} F_{\rm EUV}}{G M_{\rm p}}, 
\label{energy-limited-evap}
\end{equation}
by assuming that all the EUV energy flux is spent to accelerate the flow. 
If evaporation is powered by energy released by magnetic reconnection,  the maximum $\dot{M_{\rm p}}$ is given by an analogous formula where the magnetic power $P_{\rm mag}$ appears instead of $\pi R_{\rm p}^{2} F_{\rm EUV}$. 

\begin{figure}
\centering{
\includegraphics[width=8cm,height=8cm,angle=0]{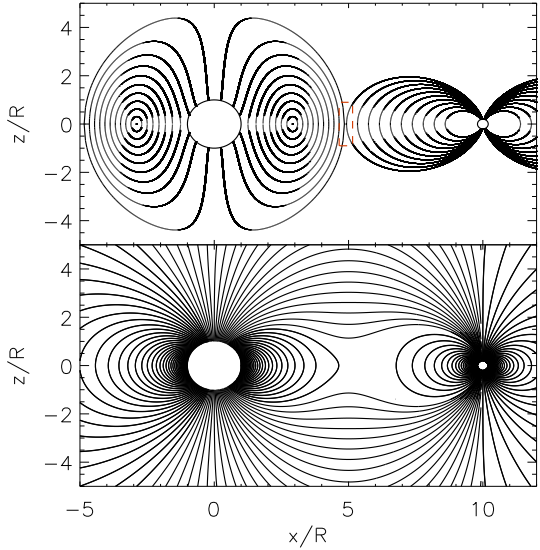}} 
\caption{Upper panel: A meridional section of the stellar and planetary magnetic field lines in a magnetohydrostatic regime in which the stellar field is force-free and the planetary field is potential. In this case the flux systems of the two bodies are topologically separated (see Sect.~\ref{coronal_model}).  The field lines of the star and the planet can release energy by reconnecting at the boundary between their magnetospheres that is highlighted with a red dashed contour. Lower panel: The magnetic field of the star-planet system in a magnetohydrostatic regime in which  the stellar field is potential. In this case, field lines interconnecting the two bodies exist. Interconnecting magnetic loops can be stressed by moving their footpoints and accumulate energy that can be successively released by a relaxation to the minimum-energy potential field configuration (see Sect.~\ref{coronal_model} for details).} 
\label{magnetic_cases}
\end{figure}

In principle, two different mechanisms can be invoked to produce a release of magnetic energy \citep[e.g., ][]{Lanza09,Lanza12}: a) reconnection between  stellar and planetary magnetic fields at the magnetospheric boundary; b) relaxation of a stressed magnetic loop interconnecting stellar and  planetary fields. 
{ Fig.~\ref{magnetic_cases} provides a sketch of the magnetic field configurations associated with these mechanisms that will be further discussed in Sect.~\ref{coronal_model}. The power made available by  mechanism a) is remarkably smaller than that of b) (cf. Sect.~\ref{star-planet_inter}). }Therefore,  to adopt a conservative approach in the present exploratory investigation, we consider only the power released by  magnetic reconnection at the boundary of the planetary magnetosphere. The  power released in the relaxation of an interconnecting loop will be considered in Sect.~\ref{strong-interaction}.

We assume that the stellar magnetic field $B_{\rm s}$ varies with  distance $r_{\rm s}$ from the centre of the star as:
\begin{equation}
B_{\rm s} = B_{0} \left( \frac{r_{\rm s}}{R_{\rm s}} \right)^{-s},
\label{stellar-field-var}
\end{equation}
where $B_{0}$ is the field at the pole of the star, $R_{\rm s}$ the star radius, and the exponent $2 \leq s \leq 3$ depends on the geometry of the field lines. Specifically, for a radially directed field, as in the case of a strong stellar wind that blows the field lines nearly radially away from the star, $s =2$, while for a potential dipole field $s=3$. Values in between these two extrema are also possible, as in the case of the non-linear force-free coronal field discussed in Sect.~3.3 of \citet{Lanza12}. The magnetic field of the planet is assumed to be dipolar with the axis perpendicular to the plane of the orbit. 

The power $P_{\rm mag}$ released by magnetic reconnection is estimated according to the model in Sect.~2.2 of \citet{Lanza12}:
\begin{equation}
P_{\rm mag} = \frac{\pi}{2\mu} B_{0}^{2} R_{\rm p}^{2} \sqrt{\frac{GM_{\rm s}}{a}} \left( \frac{B_{0}}{B_{\rm p}} \right)^{-2/3} \left( \frac{a}{R_{\rm s}} \right)^{-4s/3},
\label{p-mag}
\end{equation}
where $\mu$ is the magnetic permeability of the plasma, $B_{\rm p}$ the planetary magnetic field at its pole,  $M_{\rm s} \gg M_{\rm p}$ the mass of the star, and $a$ the semimajor axis of the orbit assumed to be circular.  

\subsection{Stellar EUV fluxes}
\label{estimateofeuvflux}

Unfortunately, photons with a wavelength shorter than 91.2~nm are strongly absorbed by  neutral hydrogen in the interstellar medium, thus only very limited information is available on the fluxes of stars in the wavelength range $10 \la \lambda \la 90$~nm that provides most of the energy for photoevaporation of planetary atmospheres.  \citet{Lecavelierdesetangs07} and other authors have extrapolated the EUV fluxes from the observed X-ray fluxes, assuming the same EUV-to-X-ray flux ratio of the Sun. An alternative is to build models of  stellar coronal emission, calibrated with solar observations and the few stellar measurements presently available, to compute fluxes in the EUV domain \citep{SanzForcadaetal11}. Although  the method of combining models and observations is likely to provide more accurate results, here we prefer to use the simpler approach of \citet{Lecavelierdesetangs07}. Specifically, we adopt the EUV fluxes given in his Sect.~3 as representative of the irradiation of stars of different spectral types in the middle of their main-sequence evolution.

\subsection{Stellar and planetary magnetic fields}
\label{stellar-fields}

The surface magnetic fields of some planet-hosting stars have been measured using spectropolarimetric techniques. For F-type stars,  poloidal fields up to $\sim 10$~G have been found in the case of HD~179949 \citep{Faresetal12} and $\tau$~Bootis \citep{Donatietal08,Faresetal09}. For G-type stars, we have a global poloidal field of $1-1.5$~G in the Sun, while fields up to $3-5$~G are observed in stars with rotation periods in the range $20-23$ days. Stars with a rotation period of $9-12$~days show fields of $\approx 40-50$~G \citep{Petitetal08}. In the active K-type star HD~189733, that has a rotation period of $\sim 12$~days,  
\citet{Moutouetal07} measured  fields up to $\sim 40$~G. Comparable or stronger fields are present at the surface of M-type dwarfs \citep[see ][ for more information]{DonatiLandstreet09}. In view of these observations and to be conservative, we adopt a stellar magnetic field intensity $B_{0} = 10, 5, 2, 10$~G for the ranges of effective temperatures $T_{\rm eff} \geq 6300$~K, $6000 \leq T_{\rm eff} < 6300$~K,  $5300 \leq T_{\rm eff}< 6000$~K, and $T_{\rm eff} < 5300$~K, respectively. They correspond to the spectral type ranges F6-F7, F8-F9, G, and K plus M of Sect.~3 of \citet{Lecavelierdesetangs07}, respectively. 

Planetary magnetic fields have not been measured yet, although a single radio flux measurement at 150~MHz suggests a field intensity  of $\approx 50$~G in HAT-P-11b to be confirmed by further observations \citep{Lecavelierdesetangsetal13}. Note that in the case of Jupiter, the field strength derived from its radio emission is $\sim 14$~G. An alternative way to estimate planetary fields is provided by the observations  of magnetospheric bow shocks  in   transiting planets. Interpreting the observations of the transit of WASP-12 in the UV domain with this model, \citet{Vidottoetal10b} estimate an upper limit to the planetary field  of $\sim 24$~G. Theoretical models computed by \citet{Christensenetal09} and \citet{ReinersChristensen10} indicate that magnetic fields between $\sim 10$ and $\sim 100$~G should be typical of planets with a mass between $\sim 1$ and $\sim 10$ Jupiter masses, with  lower values characteristic of older planets because the internal heat flux that powers their magnetohydrodynamic dynamos decreases with age. { \citet{BatyginStevenson10} find similar fields by scaling according to the planet Elsasser number computed by assuming tidal locking between its rotation and orbital revolution \citep[see also][ for a similar approach in the case of solar system planets]{RudigerHollerbach04}.}
Therefore, it seems reasonable to adopt  $B_{\rm p} = 10$~G for all the massive planets considered in the present investigation. 

\subsection{Comparison of the input powers}
\label{powers-comp}

\begin{figure}
\centering{
\includegraphics[width=8cm,height=10cm,angle=0]{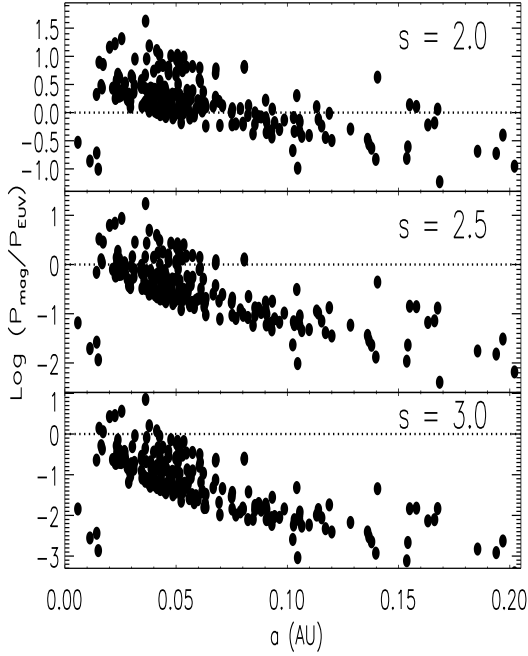}} 
\caption{Upper panel: The ratio $P_{\rm mag}/P_{\rm EUV}$ vs. the orbital semimajor axis $a$ for the known  transiting planets computed for a stellar radial magnetic field ($s=2$). The value of $P_{\rm mag}$  given by Eq.~(\ref{p-mag}) has been multiplied by $1/6$ (see the text). The dotted horizontal line corresponds to $P_{\rm mag} = P_{\rm EUV}$. Middle panel: the same as the upper panel, but for $s=2.5$ corresponding to a non-linear force-free stellar field. Lower panel: the same as the upper panel, but for $s=3$, corresponding to a potential dipole stellar field. }
\label{pmagvspeuv}
\end{figure}

The ratio $P_{\rm mag}/P_{\rm EUV}$, as obtained from Eqs.~(\ref{p-mag}) and (\ref{p-euv}), is plotted vs. the orbital semimajor axis in Fig.~\ref{pmagvspeuv} for the transiting planets known on 25 March 2013 whose parameters have been extracted from the database in www.exoplanets.org. We conservatively assume that only 30 percent of the power given by Eq.~(\ref{p-mag}) is converted into accelerated electrons and that only half of them reach the planetary atmosphere, i.e., only $1/6$ of the reconnection power is available for evaporation (cf. Sect.~\ref{electron-acceleration}).
Note that the ratio $P_{\rm mag}/P_{\rm EUV}$ does not depend on the radius of the planet. We consider three stellar magnetic field geometries as parametrized by $s=2, 2.5$, and $3$. Considering the 200 planets within 0.1~AU, the magnetic reconnection power exceeds the EUV power in 
81.5, 22.5, and 4 percent of the cases for $s=2, 2.5$, and 3, respectively. Considering the 115 planets with $a \leq 0.05$~AU, 
the percentages rise to 92.1, 26.1, and 6.9, respectively. { The four planets in the lower left corners of the plots are the three planets of Kepler-42 and GJ~1214b that orbit low-mass stars for which the ratio $a/R_{\rm s}$ is significantly greater than in the case of the other close-in planets thus making their $P_{\rm mag}$ remarkably small.}

The mean values of the ratio $P_{\rm mag}/P_{\rm EUV}$ are $\sim 2.5$, $\sim 0.7$, and $\sim 0.2$ for $s=2, 2.5$, and $3$, respectively. Therefore, the magnetic-induced evaporation should not significantly modify the predictions by \citet{Lecavelierdesetangs07} on the timescale required for a complete evaporation of a given planet, except possibly for $s=2$. 
However, in a planet endowed with a significant magnetic field, evaporation by EUV irradiation is prevented by the closed field lines in the low-latitude region and only at high latitudes (say, $ \ga 60^{\circ}$) the flow is allowed to leave the planet along open field lines \citep[cf. ][]{Adams11,Trammelletal11}. Therefore, the EUV-induced evaporation is  significantly less efficient than assumed by \citet{Lecavelierdesetangs07}, if planetary magnetic fields are not negligible. The magnetic-induced evaporation may compensate for this reduction if $s \la 2.5$. 

We conclude that the power released by magnetic star-planet interaction  can play a relevant role in the evaporation of several close-in planets and is therefore worth of a detailed investigation. A first step in this direction is undertaken in the next Sections of this work. 

\section{Model}
\label{model}

\subsection{The stellar corona}
\label{coronal_model}

When the kinetic energy,  gravity and thermal pressure $p$ of the plasma are much smaller than the magnetic pressure $B^2/2\mu$,  the field rules the dynamics and the energy balance of the plasma. In this case, a magnetohydrostatic configuration is characterized by the vanishing of the Lorentz force, i.e.,  ${\vec J} \times {\vec B} =0$, where $\vec J = \mu^{-1} \nabla \times {\vec B}$ is the current density. In other words, $\nabla \times {\vec B}  =\alpha {\vec B}$, where the force free-parameter $\alpha$ is constant along magnetic field lines, as immediately follows by taking the divergence of both sides of the equation \citep[see, e.g., ][ for more details]{Lanza12}. 

A potential magnetic field has $\alpha =0 $ along all its field lines and represents the minimum energy state compatible with the boundary conditions. The stellar field is not in general in a potential state because  the magnetic stresses at the stellar photosphere store energy in the field itself making it  deviate from its minimum-energy state. The relaxation of the field to its  potential state is generally prevented by the conservation of magnetic helicity \citep[see, ][ for details]{Lanza09,Lanza12}. On the other hand,  the magnetic field of the planet is in a potential state ($\alpha= 0$) because currents are generally not allowed to circulate in the nearly neutral planetary atmosphere. Therefore, in the considered magnetohydrostatic regime,  a magnetic field line cannot connect the planet with the stellar surface, except when the stellar field is potential, because $\alpha$ is constant along field lines. 

If the stellar field is non-potential, the flux systems of the star and the planet are topologically separated and their magnetic field lines come into contact on the boundary of the planetary magnetosphere { (cf. Fig.~\ref{magnetic_cases}, upper panel)}. In general, the field lines belonging to the two systems are not parallel on the magnetospheric boundary and the orbital motion of the planet pushes them into contact on one side of the surface forcing continuous  magnetic  reconnection. Considering that the Alfven velocity in the stellar corona is generally one order of magnitude larger than the orbital velocity of the planet and that the ratio $\beta \equiv 2\mu p/B^{2} \ll 1$ \citep{Lanza08,Lanza09,Lanza12}, the magnetic pressures of the stellar and planetary fields are in equilibrium at each point of the boundary, i.e., $B_{\rm s}^{2}(\vec r_{\rm s}) = B_{\rm p}^{2} (\vec r_{\rm p})$, where $\vec B_{\rm s}(\vec r_{\rm s})$ is the stellar field at  position $\vec r_{\rm s}$ with respect to the barycentre of the star, and $\vec B_{\rm p} (\vec r_{\rm p})$ the planetary field at  position $\vec r_{\rm p}$ with respect to the barycentre of the planet. 

When the magnetic helicity of the stellar field is negligible, the coronal field can relax to its minimum-energy potential state and in this case there are  magnetic field lines interconnecting the surface of the star with the planet { (cf. Fig.~\ref{magnetic_cases}, lower panel)}. Close to the planet, we  approximate the potential field as in Sect.~3 of \citet{Adams11} because the radii of the star and the planet are much smaller than their separation \citep[cf. also ][]{Adamsetal11}. 
 
Considering a spherical polar reference frame $(O, r_{\rm s}, \theta, \varphi)$ with the origin $O$ at the barycentre of the star and the polar axis $\hat{z}$ along the stellar spin axis, a stationary flow of the coronal plasma is ruled by the equation:
\begin{equation}
\rho ({\vec v} \cdot \nabla ) {\vec v} = -\nabla p + \rho \nabla \Phi + {\vec J} \times {\vec B}, 
\label{equileq}
\end{equation}
where $\rho$ is the density, $\vec v$ the velocity of the plasma, and $\Phi = GM_{\rm s}/r_{s} + GM_{\rm p}/r_{\rm p} + (1/2) \Omega^{2} r_{\rm s}^{2} \sin^{2} \theta$ the total gravitational plus centrifugal potential, with  $\Omega$ the angular velocity of rotation of the star, and $\theta$ the colatitude measured from its North pole. In the regime $\beta \ll 1$,  the plasma flows along magnetic field lines, that is $\vec v = v \hat{\vec s}$, where $\hat{\vec s}$ is a unit vector in the direction of the magnetic field, i.e., $\vec B = B \hat{\vec s}$. Since thermal conduction is very efficient along  field lines at temperatures of the order of $\sim 10^{6}$~K \citep{Priest84},  we assume that the coronal temperature $T_{\rm c}$ is constant along a given field line, i.e., $\partial T_{\rm c} / \partial s = 0$, and that the plasma follows the ideal gas law: $p= (\tilde{R}/\tilde{\mu}) \rho T_{\rm c}$, where $\tilde{R}$ is the gas constant and $\tilde{\mu}$ the mean molecular weight. Under these hypotheses, Eq.~(\ref{equileq}) can be integrated along a given field line by considering that $ ({\vec v} \cdot \nabla) {\vec v} = \nabla (v^{2}/2) + (\nabla \times \vec v) \times {\vec v}$, and gives:
\begin{equation}
\rho = \rho_{0} \exp \left\{ \frac{\tilde{\mu}}{\tilde{R}T_{\rm c}} \left[ -\frac{1}{2} v^{2} + (\Phi - \Phi_{0}) \right] \right\},
\label{density-eq}
\end{equation}
with the potential $\Phi$ given by:
\begin{equation} 
\Phi = G \frac{M_{\rm s}}{r_{\rm s}} \left[ 1 + \epsilon_{\rm rot} \left(\frac{r_{\rm s}}{R_{\rm s}} \right)^{3} \right] + G \frac{M_{\rm p}}{r_{\rm p}}, 
\end{equation}
where $\rho_{0}$ is the coronal density at the footpoint of the field line, $\epsilon_{\rm rot} \equiv \Omega^{2} R_{\rm s}^{3}/(2 G M_{\rm s})$ a measure of the ratio of the centrifugal to the gravitational potential on the equator of the star, and $\Phi_{0}$ the potential at the footpoint of the field line on the stellar surface. Eq.~(\ref{density-eq}) assumes that the velocity vanishes at the footpoint of the field line.

\subsection{Interaction between  stellar and planetary fields}
\label{star-planet_inter}

When the stellar field is in a force-free non-potential state, it interacts with the planetary field at the boundary of the planetary magnetosphere that is the surface of discontinuity between the two fields as discussed in Sect.~\ref{coronal_model}. \citet{Lanza09,Lanza12} has computed the power  released by magnetic reconnection at the magnetospheric boundary and its dependence on stellar and planetary parameters considering different models for the stellar coronal field. The maximum dissipated power is found to range from $10^{17}$ to $10^{19}$~W, even considering rather extreme values of the parameters. Such values are insufficient by at least one order of magnitude to account for the power radiated by the chromospheric hot spots observed by \citet{Shkolniketal05,Shkolniketal08} and 
\citet{Gurdemiretal12}. 

In the present approach, Alfven waves have been filtered out by assuming that the Alfven velocity is much larger than the orbital velocity of the planet. However, even without this assumption, it is not possible to account for powers of the order of $10^{20}-10^{21}$~W, as found by  \citet{Sauretal13} by considering the energy fluxes in the so-called Alfven wing model.

To account for greater powers, we consider a  magnetic loop interconnecting the surface of the star with the planet that is steadily stressed by the orbital motion of the planet. The accumulation of energy drives the field out of its initial potential state as in the case of a solar magnetic arcade that is stressed by  shear motions at its footpoints. At a given point, the arcade losses equilibrium and the field erupts producing a  flare while the magnetic helicity is taken away by a coronal mass ejection allowing the field to relax again to a nearly potential state  \citep[e. g., ][]{Flyeretal04,Zhangetal06,ZhangFlyer08}. The same process is expected to occur in the case of an interconnecting star-planet loop. 
The effectiveness of the eruption process in taking away magnetic helicity is crucial in this case because, if helicity accumulates in the field, it will lead to a separation of the stellar and planetary flux systems, i.e., the disappearance of the interconnecting loop  (cf. Sect.~\ref{coronal_model}).

Assuming a sequence of cycles consisting of stress accumulations and eruptions, the average energy dissipation rate in the interconnecting loop must be equal to the average flux of the Poynting vector across its base. In other words, the mean available power is:
\begin{equation}
P = \frac{2\pi}{\mu} f_{\rm AP} R_{\rm p}^{2} |{\vec E} \times {\vec B}|,  
\end{equation}
where $f_{\rm AP}$ is the fraction of the planetary hemisphere $2 \pi R_{\rm p}^{2}$ crossed by the field lines of the interconnecting loop, $\vec E = -{\vec v_{0}} \times {\vec B_{\rm p}}$ the electric field, $\vec v_{0}$  the relative orbital velocity, and  $\vec B_{\rm p}$ the magnetic field at the base of the loop, located on the surface of the planet. In terms of the surface field of the planet, we have:
\begin{equation}
P \simeq \frac{2\pi }{\mu} f_{\rm AP} R_{\rm p}^{2} B_{\rm p}^{2} v_{0}. 
\label{maximum_power}
\end{equation}
The fraction of the planetary surface magnetically connected to the stellar field is given by Eq.~(27) of \citet{Adams11}: 
\begin{equation}
f_{\rm AP} \equiv 1-\left( 1 - \frac{3 \zeta^{1/3}}{2 +\zeta} \right)^{1/2}, 
\label{adamsfap}
\end{equation}
 where $\zeta \equiv B_{\rm s}(a)/B_{\rm p}$ is the ratio of the magnetic field  of the star to that of the planet at the distance $a$ of the planet (cf. Eq.~\ref{stellar-field-var}). Note that our parameter $\zeta$ is denoted as $\beta$ in \citet{Adams11},  the symbol being changed here to avoid  confusion with the plasma $\beta$. { Eq.~(\ref{maximum_power}) is valid when the timescale for the field relaxation to the potential state $\tau_{\rm rel}$ is comparable with or shorter than the timescale for magnetic energy buildup by the stress produced by the orbital motion of the planet $\tau_{\rm acc}$. The typical value of $\tau_{\rm rel}$ is comparable with the Alfven transit time along the interconnecting loop and is of the order of $\sim 10^{3}$~s for close-in planets \citep[cf., e.g., the final part of Sect.~2.3 and Sect.~3.1 of ][]{Lanza12}, while $\tau_{\rm acc} \sim 2R_{\rm p}/v_{0}$ is the timescale for the orbiting planet to cross the base of the interconnecting loop and is of the order of $10^{3}-10^{4}$~s for typical $R_{\rm p} \sim 10^{8}$~m and $ v_{0} \sim 10^{4}-10^{5} $~m~s$^{-1}$, thus satisfying the requirement $\tau_{\rm rel} \leq \tau_{\rm acc}$. }

For a  field strength $B_{0} \sim B_{\rm p} \sim 10$~G, a relative velocity $v_{0} \sim 10^{4}-10^{5}$~m~s$^{-1}$, a planetary radius $R_{\rm p} \sim 10^{8}$~m, and $f_{\rm AP} \sim 0.1 - 0.2$, we find $P \sim 10^{20}-10^{21}$~W, that is sufficient to explain the power radiated by chromospheric hot spots associated with HJs. That given by Eq.~(\ref{maximum_power}) is indeed the maximum available power that one can expect from magnetic star-planet interaction given the hypotheses adopted here. { The energy released in the  interconnecting loop is ultimately provided by the orbital motion of the planet, but its dynamical effect on the orbit is negligible in comparison with, e.g., tidal interactions because such a kind of loop is present only for a limited fraction of the time  \citep[$\approx 10-20$ percent; cf., e.g., Sect.~4.2 of ][]{Lanza11}.}

The constant phase lag between the planet and the hot spot \citep{Shkolniketal05} can be explained by assuming that the coronal field is initially in an axisymmetric  twisted force-free regime \citep[][]{Lanza08,Lanza09} before reconnecting with the planetary field. During the loss of equilibrium occurring in the reconnection phase, the interconnecting field gets rid of its magnetic helicity by erupting twisted flux tubes as in solar coronal mass ejections, thus relaxing to a nearly potential configuration. The footpoints of the interconnecting loop on the stellar surface cannot move because the field is frozen to the photospheric plasma. In this way, the subsequent stress is applied to a nearly potential loop with the initial phase lag between its stellar and planetary footpoints  almost unchanged.  If the planet crosses a sequence of loops of an unperturbed force-free axisymmetric field with an interaction timescale short in comparison with its orbital period,  an approximately constant phase lag between the planet and the chromospheric hot spot is expected. If the stellar field has a significant non-axisymmetric component, the time dependence of the energy release becomes remarkably complicated and neither a constant phase lag nor a simple dependence of the irradiated power on the orbital phase is expected \citep[cf., ][ for details]{Lanza12}. 

The energy dissipation in a stressed interconnecting loop is not localized at the boundary of the planetary magnetosphere, as in the case of the interaction of  the separated star-planet flux systems. Applying a large-scale shearing motion   to the footoints of a solar coronal loop  whose magnetic field lines are steadily intermixed by turbulent photospheric motions, \citet{JanseLow09,JanseLow10} find that a spatially dense distribution of current sheets is produced extending over a large fraction of the loop volume. Given the similarity with our case, we conclude that the dissipation of magnetic energy occurs over most of the volume of the interconnecting loop. The loop sections having the largest field intensities, i.e., those close to the stellar and the planetary surfaces, are expected to have the highest energy release rates per unit volume. 

\subsection{Electron acceleration}
\label{electron-acceleration}

Since the energy released by magnetic reconnection is mainly transported into the planetary atmosphere by accelerated electrons, we now
consider their acceleration process in some detail. 
Electron acceleration is characteristic of solar flares and reveals itself through the emission of pulses of hard X-rays from the footpoints of flaring loops where the particles impinge upon the dense chromospheric plasma and emit hard X-ray bremsstrahlung radiation. The acceleration  occurs at the reconnection sites inside  flaring loops or magnetic arcades generally located up in the solar corona \citep[see, e.g., ][ for a review]{Aschwanden02}. In a large flare, the total number of electrons accelerated to energies of 20 keV or higher is of the order of $10^{38}-10^{39}$ and the timescale of acceleration is as short as $\approx 300$~ms for an individual pulse \citep{Milleretal96}. The total energy converted into  accelerated electrons is a significant fraction of the total flare energy generally ranging from $\approx 30$ to $60$ percent  \citep[e. g., ][]{Holman05,MannWarmuth11}. On the other hand, the energy of other particles (protons, ions) is generally lower or comparable with that of the electrons \citep[see ][ Sect.~7.4]{Aschwanden02}, therefore we shall consider only electrons for simplicity.  

Several different acceleration mechanisms have been proposed to occur in the reconnection sites of solar flares \citep[cf., e. g., ][]{Milleretal96,Aschwanden02,Petrosian12}. Here we adopt a simple heuristic approach to relate  the spectrum of the accelerated electrons to the average magnetic field and  particle density in the reconnection region, inspired by the discussion in Sect.~4.1.1 of \citet{Aschwanden02}. The density spectrum of the accelerated electrons is denoted by ${\cal S}(E_{0})$, where $ {\cal S} (E_{0}) d E_{0}$ is the number of particles with kinetic energy between $E_{0}$ and $E_{0} + d E_{0}$ per unit volume. We assume that the spectrum has the form: 
\begin{equation}
{\cal S} (E_{0}) = \left\{ 
\begin{array}{l}
A \; \mbox{    if $E_{0} < E_{\rm c}$ }, \\
A (E_{0}/E_{\rm c})^{-\delta} \; \mbox{if $E_{0} \geq E_{\rm c}$},
\end{array}
\right.
\label{energy-spectrum}
\end{equation}
where $E_{\rm c}$ is the so-called cutoff energy, $\delta$  the index of the power-law distribution for energies beyond the cutoff, and $A$ a normalization constant (see below). { A positive and finite value of the total electron energy per unit volume requires $\delta > 2$ (cf. Eq.~\ref{total-electron-energy}).} The power-law dependence of the energy spectrum follows from the spectrum of hard X-rays observed during the impulsive phase of  solar flares \citep[see, e.g., ][]{Holmanetal11}. The energy cutoff $E_{\rm c}$ is generally estimated to fall between $15$ and $30$~keV, but a unbiased  determination of its range is difficult because  the thermal X-ray component emitted by the very hot plasma produced during a flare  hides the change in slope of the X-ray bremsstrahlung spectrum when the cutoff is at low energies. The power-law index $\delta$ usually ranges between 4 and 6. The energy distribution below the cutoff is assumed to be flat because a sharp or peaked cutoff or a positive slope would lead to instabilities that rapidly re-establish a flat  distribution, as discussed in Sect.~6 of \citet{Holmanetal11}. 

If we denote by $n_{\rm T}$ the number density of the accelerated electrons, we have:
\begin{equation}
n_{\rm T} = \int_{0}^{\infty} {\cal S}(E_{0}) dE_{0},
\label{aconstant_norm}
\end{equation}
yielding $A=(n_{\rm T}/E_{\rm c})(\delta-1)/\delta$.  
The total energy per unit volume is:
\begin{equation}
E_{\rm T} = \int_{0}^{\infty} E_{0} {\cal S}(E_{0}) dE_{0} = \frac{1}{2} \left( \frac{\delta-1}{\delta -2} \right) n_{\rm T} E_{\rm c}. 
\label{total-electron-energy}
\end{equation}
The source of energy for electron acceleration is the coronal magnetic field. On average, some fraction $f$ of the magnetic energy per unit volume $B^{2}/2 \mu$ is converted into $n_{\rm T}$ accelerated electrons with a total energy $E_{\rm T}$. If $g$ is the fraction of accelerated electrons, i.e., $n_{\rm T} = g n_{\rm e}$, where $n_{\rm e}$ is the electron number density, we have:
\begin{equation}
\frac{1}{2}  \left( \frac{\delta-1}{\delta -2} \right) g n_{\rm e} E_{\rm c} = f \frac{B^{2}}{2\mu}.
\label{acc-efficiency}
\end{equation}
Assuming in a typical solar flare $B \sim 100$~G, $n_{\rm e} \sim 10^{16}$~m$^{-3}$ \citep{Milleretal96}, $E_{\rm c} \sim 25$~keV, and $f \sim 0.3$, we obtain from Eq.~(\ref{acc-efficiency}), $g \sim 0.1-0.3$, that is a significant fraction of the coronal plasma electrons is accelerated to suprathermal energies in a solar flare. Conversely, when the fraction of accelerated electrons $g$, the energy  conversion fraction $f$, the power-law index $\delta$, the electron density $n_{\rm e}$, and the field strength $B$ are given, we can use Eq.~(\ref{acc-efficiency}) to estimate the cutoff energy $E_{\rm c}$. This approach will be applied in Sects.~\ref{energy-dissipation} and~\ref{results} to estimate $E_{\rm c}$ for the electrons accelerated  at the boundary of the planetary magnetosphere.  The magnetic field intensity $B$ is given by Eq.~(\ref{stellar-field-var}), while the number density $n_{\rm e}$ follows from Eq.~(\ref{density-eq}) assuming that the plasma is fully ionized.

\subsection{Energy dissipation in the outer planetary atmosphere}
\label{energy-dissipation}

After their acceleration,  high-energy electrons travel along  magnetic field lines reaching the stellar chromosphere at one footpoint and the planetary atmosphere at the other footpoint. The energization of the stellar atmosphere has been considered by, e.g., \citet{GuSuzuki09}, thus we focus on the energy dissipated into the planetary atmosphere. 
We follow \citet{Holmanetal11} to determine the energy lost  by the electrons during their travel through the atmospheric layers. The atmosphere is considered to be isothermal with a temperature $T$ (see Sect.~\ref{atmosphere} for  justification) and the tidal gravitational potential of the star is neglected. For simplicity,  we assume a constant ionization fraction $x \equiv n_{\rm p}/n$, where $n$ is the total number density of protons, including both free protons of density $n_{\rm p}$ and those bound into hydrogen atoms of density $n_{\rm H}$, i.e., $n=n_{\rm p} + n_{\rm H}$. The layers from which the planetary evaporation starts have a typical density $ n \approx 10^{15}-10^{17}$~m$^{-3}$ and a temperature of $\approx 10^{4}$~K \citep{Murray-Clayetal09,Trammelletal11,Adams11}, giving a pressure of $10^{-9}-10^{-7}$~bar. In those layers, the chemical composition is dominated by hydrogen \citep[cf., e.g., ][]{Venotetal12}. Therefore, it is justified to assume an atmosphere of pure hydrogen.
Our model applies to the high-latitude zones of the planet where the footpoints of the interacting magnetic field lines are located and the magnetic field can be considered to be vertically directed. The speed of the evaporation flow is generally smaller than the  sound speed in the layers where most of the energy is deposited, therefore  the stratification is not remarkably different from hydrostatic. In other words, we assume that the number density of  protons in the atmosphere varies with the height $z$ from its base layer as: $n(z) = n_{\rm p0} \exp(-z/H)$, where $n_{\rm p0}$ is the base density and $H = \tilde{R} T (1+x)/g $ is the atmosphere scale height, with $g=GM_{\rm p}/R^{2}_{\rm p}$ the gravitational acceleration at the surface of the planet and the mean molecular weight for a pure hydrogen partially-ionized atmosphere $\tilde{\mu} = 1/(1+x)$. 

The loss of energy $dE$ experienced by a single electron of energy $E$ crossing a layer between the heights $z$ and $z+dz$ is \citep[see ][ Sect. 4.2]{Brown73,Holmanetal11}:
\begin{equation}
\frac{dE}{dz} = - \frac{K^{\prime}}{E} (\lambda + x) n (z),
\label{energy_loss}
\end{equation}
where $K^{\prime} = 2 \pi c^{4}e^{4} \Lambda$, with $c$ the speed of light, $e$ the electron charge, $\Lambda \equiv \Lambda_{\rm ee} - \Lambda_{\rm eH}$, $\lambda \equiv \Lambda_{\rm eH}/\Lambda$, with $\Lambda_{\rm ee}$ and $\Lambda_{\rm eH}$ the Coulomb logarithms for electron-electron and  electron-hydrogen atom collisions, respectively. Since $\Lambda_{\rm ee} \simeq 20$ and 
$\Lambda_{\rm eH} \simeq 7.1$ are almost constant in the temperature and density regime of interest, $\Lambda \simeq 12.9$ and $\lambda \simeq 0.55$. Therefore, the constant $K^{\prime} \simeq 1.87 \times 10^{-22} (\Lambda/12.9)$ keV$^{2}$ m$^{2}$. Assuming that the ionization fraction $x$ is constant, 
Eq.~(\ref{energy_loss}) can be integrated to give the energy of an electron at height $z$:
\begin{equation}
E(z) = \sqrt{E_{0}^{2} - 2 K^{\prime \prime} N_{\rm c}(z)}, 
\label{energy-height-z}
\end{equation}
where $E_{0}$ is the energy of the particle when it leaves the acceleration site, $K^{\prime \prime} \equiv (\lambda + x) K^{\prime}$, and $N_{\rm c} (z) = \int_{z}^{\infty} n(z^{\prime}) d z^{\prime} $ the column density of the atmosphere down to the level $z$. If ${\cal F} (E_{0}) dE_{0}$, is the number of electrons with initial energy between $E_{0}$ and $E_{0} + d E_{0}$ that goes through the unit area on top of the atmosphere per unit time (electron flux density), the total power dissipated in the layer between $z$ and $z +dz$ per unit area is: 
\begin{equation}
\frac{dP}{dz} =  - K^{\prime \prime} n (z) \int_{E_{\rm s} (z)}^{\infty} \frac{{\cal F}_{0}(E_{0})}{\sqrt{E_{0}^{2}- E_{\rm s}^{2}(z)}} \, dE_{0} \equiv - Q_{\rm c}(z) n(z),
\label{dissipated-power}
\end{equation}
where $E_{\rm s}(z) \equiv \sqrt{2 K^{\prime \prime } N_{\rm c}(z)}$ is the minimum energy required for an electron to reach the level $z$ without being stopped by the energy losses in the upper layers of the atmosphere, and $Q_{\rm c}$ the energy loss per  proton.  Assuming that there is no significant energy loss for the electrons during the travel from their acceleration site to the limit of the atmosphere \citep[an assumption supported by solar flare observations; cf., e.g., ][ Sect.~6]{Aschwanden02}, the flux density spectrum at the limit of the atmosphere is ${\cal F}_{0}(E_{0}) = {\cal S} (E_{0}) v(E_{0})$, where $v(E_{0}) = \sqrt{2 E_{0}/m_{\rm e}}$ is the velocity of the electrons of energy $E_{0}$ and mass $m_{\rm e}$. From  Eq.~(\ref{energy-spectrum}), we obtain:
\begin{equation}
{\cal F}_{0}(E_{0})= \sqrt{\frac{2}{m_{\rm e}}} n_{\rm T} E_{\rm c}^{-1/2} \left( \frac{\delta -1}{\delta} \right) \left\{ 
\begin{array}{l}
(E_{0}/E_{\rm c})^{1/2} \; \;\; \mbox{if $E_{0} < E_{\rm c}$ }, \\
(E_{0}/E_{\rm c})^{(1/2)-\delta} \;  \; \; \mbox{if $E_{0} \geq E_{\rm c}$}. 
\end{array}
\right.
\label{energy-flux-spectrum}
\end{equation}
On the other hand, the energy flux density of  accelerated electrons at the limit of the atmosphere is: $E_{0} {\cal F}_{0}(E_{0})$. Denoting by $F_{0}$ the total energy flux carried by accelerated electrons (energy reaching the top of the atmosphere per unit surface and time) and integrating over the energy $E_{0}$, we find:
\begin{equation}
F_{0} = \frac{4 (\delta -1)}{5(2\delta - 5)} \sqrt{\frac{2}{m_{\rm e}}} n_{\rm T} E_{\rm c}^{3/2}, 
\label{total-energy-flux}
\end{equation}
that gives the number density of accelerated electrons $n_{\rm T}$ from $F_{0}$, the cutoff energy $E_{\rm c}$, and the slope of the spectrum $\delta$. 

Substituting Eq.~(\ref{energy-flux-spectrum}) into Eq.~(\ref{dissipated-power}), we can compute the energy dissipated by accelerated electrons into a given layer of the atmosphere. We express the integrand in terms of the non-dimensional quantities $w \equiv E_{0}/E_{\rm c}$ and $\alpha_{\rm e} \equiv E_{\rm s}(z)/E_{\rm c}$ and integrate by parts to eliminate the numerical divergence of the integrand for $E=E_{\rm s}$,  obtaining:
\begin{eqnarray}
Q_{\rm c} (z)  & =  &  K^{\prime \prime} S_{0} \left\{ \frac{1}{2} \int_{\alpha_{\rm e}}^{1} w^{-3/2} \sqrt{w^{2}-\alpha_{\rm e}^{2}} \, dw \, + \right.  \label{q-expression} \\
      &  & \left. \left( \delta + \frac{1}{2}\right)  \int_{1}^{\infty} x^{-3/2-\delta} \sqrt{w^{2} - \alpha_{\rm e}^{2}}\, dw \right\} \mbox{ for $\alpha_{\rm e} < 1$, or } \nonumber \\
      & = & K^{\prime \prime} S_{0} \left( \delta + \frac{1}{2} \right) \int_{\alpha_{\rm e}}^{\infty} w^{-3/2 -\delta} \sqrt{w^{2}- \alpha_{\rm e}^{2}} \, dw \mbox{ for $\alpha_{\rm e} \geq 1$}, \nonumber
\end{eqnarray}
where $S_{0} \equiv \sqrt{2/m_{\rm e}} n_{\rm T} E_{\rm c}^{-1/2} (\delta-1)/\delta $.  These expressions  can be integrated numerically by truncating the upper limit of the improper integrals to some value $w_{\rm max} \gg \max (1, \alpha_{\rm e}) $ since the error introduced by this approximation does not exceed $(w_{\rm max})^{-\delta}/\delta$. 

{ When $s=2$,  in the layer where the planetary evaporation flow is launched $\alpha_{\rm e} \ll 1$ in several cases.} In this limit, we develop the square root factors in the above integrals by  Newton's binomial expansion and integrate the corresponding power series. Limiting ourselves to the terms up to the second order in $\alpha_{\rm e}$, we find:
\begin{equation}
Q_{\rm c} (z) = K^{\prime \prime} S_{0} \left[ \left(\frac{4\delta}{2\delta -1} - \frac{197}{168}\alpha_{\rm e}^{1/2} \right) -  \frac{2 \delta}{6\delta +9}  \alpha_{\rm e}^{2} \right] + O(\alpha_{\rm e}^{4})
\label{q-limit}
\end{equation}
For $\delta =4-6$ and $\alpha_{\rm e} \ll 1$, the dissipation rate is dominated by the first two terms, thus $Q_{\rm c}$ is approximately constant because $\alpha_{\rm e}^{1/2} \propto N_{\rm c}(z)^{1/4}$
has a weak dependence on the column density in this limit. 

{ The above treatment is adequate as far as the energy of most of the electrons is non-relativistic, i.e., $E_{\rm c} \la 0.2 m_{\rm e} c^{2} \sim 100$~keV, where $m_{\rm e} c^{2}$ is the electron rest energy \citep[e.g., ][]{Emslie78,LeachPetrosian81}. However, when $s=2$ in $\sim 7$ percent of the cases $E_{\rm c}$  exceeds the above  limit  requiring a relativistic treatment of the collisional energy loss and ionization rates (cf. Sect.~\ref{results}). This fraction increases if we consider parameters for the stellar corona that lead to a lower particle density or a stronger magnetic field in the reconnection region than assumed in Sect.~\ref{results}. Therefore, we treat the case of relativistic electrons  in Appendix~\ref{relativity}.}

\subsection{Ionization and thermal balance of the outer atmosphere}
\label{atmosphere}

We now consider the response of the outer planetary atmosphere to the accelerated electrons and the stellar ultraviolet radiation. The ionization fraction $x= n_{\rm p}/n$ is given in equilibrium by a balance between ionization (both radiative and collisional) and recombination processes. The plasma is electrically neutral, so the number densities of the electrons and protons are equal, i.e., $n_{\rm e}=n_{\rm p}$. Adopting the approximations introduced in Sect.~8 of \citet{Trammelletal11}, the ionization balance can be written as:
\begin{eqnarray}
n_{\rm H}(J + C) & = & \alpha_{\rm R} n_{\rm e} n_{\rm p}= \alpha_{\rm R} n_{\rm p}^{2}, \;  \; \; \mbox{or, } \nonumber \\
(J+C) (1-x)  & = &\alpha_{\rm R} n x^{2},   
\label{ioniz_balance}
\end{eqnarray}
where $\alpha_{\rm R} \simeq 2.6 \times 10^{-19} (T/10^{4} K)^{-0.8}$~m$^{3}$ s$^{-1}$ is the recombination rate and $J = \int_{\nu_{0}}^{\infty} d \nu f({\nu}) \sigma_{\rm pi} (\nu) \exp [-N_{\rm cH} \sigma_{\rm pi}(\nu)]$   the photoionization rate per hydrogen atom, with $N_{\rm cH} = (1-x) N_{\rm c}$ the atomic hydrogen column density  from the given point to the star, $\nu$ the frequency of the ionizing photon, $f({\nu})$ the photon flux per unit frequency interval, simply scaled from that of the Sun \citep{Claireetal12}, $\sigma_{\rm pi}(\nu) = \sigma_{0} (\nu/\nu_{0})^{-3}$ the radiative ionization cross section of the hydrogen, $h \nu_{0}=E_{\rm i}=13.6$~eV the ionization potential, and $\sigma_{0} = 6.3 \times 10^{-22}$~m$^{2}$ the cross section at the ionization threshold; $C =  \pi a_{0}^{2} \int_{E_{\rm i}}^{\infty} {\cal C}(E) {\cal S}(E) v(E)  dE  = \pi a_{0}^{2} \int_{E_{\rm i}}^{\infty} {\cal C}(E) \tilde{{\cal F}}(E) dE$, is the ionization rate per unit hydrogen atom by electron collisions, where $a_{0}=5.3 \times 10^{-11}$~m is the Bohr radius, ${\cal C}(E)$ the ratio between the collisional ionization cross-section and the area corresponding to the Bohr radius, ${\cal S}(E)$ the number density of the incident electrons of energy $E$,  $v(E)$ their velocity, and $\tilde{{\cal F}}(E)$ their flux density. It can be computed from the flux density at the limit of the atmosphere as given by Eq.~(\ref{energy-flux-spectrum}) taking into account the energy losses as specified by Eq.~(\ref{energy-height-z}), i.e.:
\begin{equation}
\tilde{{\cal F}} (E) =   \frac{E}{\sqrt{E^{2} + 2 K^{\prime \prime} N_{\rm c}(z)}} \, {\cal F} \left( \sqrt{E^{2} + 2 K^{\prime \prime} N_{\rm c}(z)} \right).
\label{energy-flux-spectrum-z}
\end{equation}
Eq.~(\ref{energy-flux-spectrum-z}) follows from the continuity of the electron flux through the atmosphere, i.e., $\tilde{{\cal F}}(E) dE = {\cal F}_{0}(E_{0}) dE_{0}$, and Eq.~(\ref{energy-height-z}). 

The dependence of the non-dimensional cross-section ${\cal C}(E)$ on the energy $E$ of the colliding electrons is given by \citet{Rudge68} and \citet{Dere07}. In our case, the energy of the electrons is generally much greater than $E_{\rm i}$ and the collisional cross section scales as ${\cal C}(E) \propto (E/E_{\rm i})^{-1} \ln (E/E_{\rm i})$ for $E \ga 10\, E_{\rm i}$; for $E_{\rm i} \leq E \la 10\, E_{\rm i}$, we assume $C$ equal to its mean value, i.e., $\sim 0.6$. Note that for $E/E_{\rm i} \sim 25$, ${\cal C} \sim 0.3$, therefore the low-energy part of the electron spectrum produces most of the collisional ionizations. 
 
Once the ionization fraction has been calculated in a given layer as a function of the incident EUV flux  $f(\nu)$ and the  accelerated electron flux  $\tilde{{\cal F}}(E)$, we can consider the energy balance. Since the atmosphere is approximately isothermal (see below) and the gas can move only along magnetic field lines, owing to the collisional coupling between ions, electrons, and neutral atoms, thermal conduction and convection can be neglected. The thermal balance in equilibrium is established between the heating sources (EUV flux and accelerated electrons) and  the Lyman-$\alpha$ radiation that is the most important contributor to the radiative losses in the considered range of temperature, i.e., $T \sim (1-2) \times 10^{4}$~K,  
 for an optically thin plasma \citep[cf. ][ and references therein]{Dereetal09}. Lyman-$\alpha$ emission is excited by  collisions of thermal electrons with hydrogen atoms. Assuming that all the energy of the photoelectrons produced by photoionization is converted into heat, the heating rate by the EUV flux per hydrogen atom is 
$Q_{\rm ph} = \int_{\nu_{0}}^{\infty} d \nu f({\nu}) \sigma_{\rm pi} (\nu) h(\nu-\nu_{0}) \exp [-N_{\rm cH} \sigma_{\rm pi}(\nu)]$. The heating per unit volume  by  accelerated electrons is given by Eqs.~(\ref{dissipated-power}) and~(\ref{q-expression}). Therefore,   the thermal balance equation is:
\begin{eqnarray}
 Q_{\rm ph} n_{\rm H} +  Q_{\rm c} n & = & L_{\alpha}(T) n_{\rm e} n_{\rm H},  \; \; \; \mbox{or} \nonumber \\
 \left[ (1-x) Q_{\rm ph} + Q_{\rm c} \right]  &  = &  x (1-x) n L_{\alpha}(T),
\label{thermal_balance}
\end{eqnarray}
where $L_{\alpha}(T) = 2.9 \times 10^{-32} (T/10^{4} K)^{-0.5} \exp(-1.184 \times 10^{5} K/T)$~J~m$^{3}$~s$^{-1}$ is the radiative loss function of the Lyman-$\alpha$ line \citep{DalgarnoMcCray72},  $n_{\rm H}=(1-x)n$, and $n_{\rm e}=n_{\rm p}=x n$. 
This expression for $L_{\alpha}$ is valid for $T \la 1.2 \times 10^{4}$~K. 
Note the exponential dependence on  temperature that makes $L_{\rm \alpha}$  increase rapidly with increasing temperature. In other words, a remarkable increase of the heating can be balanced by a modest increase of the temperature of the emitting plasma. On the other hand, an increase of the temperature increases the ionization, thus reducing the Lyman-$\alpha$ emissivity for $T> (1-2) \times 10^{4}$~K. The interplay between these effects makes the Lyman-$\alpha$ radiation a sort of thermostat that keeps the temperature of the upper atmospheric layers approximately constant at $\approx 10^{4}$~K for a wide range of heating rates \citep[cf., e.g., Fig.~3 in ][]{CoxTucker69}.  

\subsection{A simple isothermal wind  model}
\label{wind-model}

An illustrative application of the theory introduced above is provided by a  model of a transonic isothermal wind, i.e.,  with a uniform temperature $T$ and ionization fraction $x$  from the launching layer of the flow to the sonic point and beyond. \citet{Adams11} has introduced such a model that we  use in this Section.  

The layer from which the wind is accelerated corresponds approximately to  the layer where the optical depth for the absorption of EUV photons by  hydrogen atoms is unity. As a matter of fact, a fully consistent model  should include the energization of deeper layers by  non-thermal electrons. This would imply the solution of the full energy equation in an optically thick plasma that makes the problem much more complicated. Therefore, we limit ourselves to the assumption that the base of the planetary wind flow is at the level $z_{0}$ where the optical depth
\begin{equation}
\tau \equiv (1-x) N_{\rm c}  \langle \sigma_{\rm EUV} \rangle = 1,
\label{taueq1}
\end{equation}
 where $\langle \sigma_{\rm EUV} \rangle = 2.0 \times 10^{-22}$~m$^{2}$ is the mean absorption cross-section for the EUV radiation. In such a way, we can apply Eq.~(\ref{thermal_balance}), valid for an optically thin plasma at Lyman-$\alpha$ wavelength. 
The column density down to the level  $z$ in an isothermal, partially ionized atmosphere is: $N_{\rm c} (z) = n_{\rm p0} H \exp (-z/H)$, where $n_{\rm p0}$ is the density at the base of the atmosphere (level $z=0$). We assume $n_{\rm p0} = 2 \times 10^{16}$~m$^{-3}$ and verify that our results are not significantly dependent on that value. 

The flux of EUV radiation producing the heating and ionization of the planetary atmosphere  has been specified in Sect.~\ref{estimateofeuvflux} according to Sect.~3 of \citet{Lecavelierdesetangs07}. The energy flux $F_{0}$ of non-thermal electrons at the boundary of the atmosphere, required to compute the electron spectrum with the theory of Sects.~\ref{electron-acceleration} and~\ref{energy-dissipation}, is given by: 
\begin{equation}
F_{0} = \frac{1}{6} \, \frac{P_{\rm mag}}{2 \pi f_{\rm AP} R_{\rm p}^{2}},
\label{magnetic-flux}
\end{equation} 
where the power $P_{\rm mag}$ released in the star-planet interaction is given by Eq.~(\ref{p-mag}) and $f_{\rm AP}$ is the fraction of the planetary surface magnetically connected to the star (cf. Eq.~\ref{adamsfap}).
  In Eq.~(\ref{magnetic-flux}) a factor  $1/6$ is included to account for an efficiency of $\sim 30$ percent in electron acceleration and to consider that only  half of the electrons stream toward the planet (cf.  Sect.~\ref{powers-comp}).  To estimate the cutoff energy $E_{\rm c}$, we apply Eq.~(\ref{acc-efficiency}), with $f=0.3$, $g=0.1$, and $\delta =6$, while Eq.~(\ref{total-energy-flux}) is applied to find $n_{\rm T}$ from $F_{0}$, thus completely specifying the electron energy spectrum. 

For given EUV and non-thermal electron fluxes, we solve iteratively for the density, the temperature, and the ionization fraction at the base level $z_{0}$. Specifically, we start by guessing $T \sim 10^{4}$~K and $x \sim 0.5$ at $z_{0}$ and compute a first guess for $z_{0}$ from Eq.~(\ref{taueq1}); then we compute the density at $z=z_{0}$ from $n(z) = n_{\rm p0} \exp(-z/H)$, solve Eq.~(\ref{ioniz_balance}) to find a refined value of the ionization fraction, and Eq.~(\ref{thermal_balance}) for a refined value of the temperature (from the Lyman-$\alpha$ radiative loss function)  that are used to estimate a better value of $z_{0}$ by Eq.~(\ref{taueq1}). By iterating this procedure, we converge to consistent values for $z_{0}$, $n(z_{0})$, $T(z_{0})$, and $x(z_{0})$ giving $\tau(z_{0})=1$. Considering the parameters of the transiting exoplanets as listed in the website www.exoplanets.org on 25 March 2013, and excluding those with  a mass below $0.05$ Jupiter masses and a semimajor axis $a > 0.15$~AU, we find that the above procedure fails to converge only in $\sim 1$~percent of the cases. 

Once the physical parameters at the base of the planetary wind are known, the mass loss rate is obtained by multiplying the non-dimensional mass loss rate $\dot{m}$ given by Eq.~(64) of \citet{Adams11} by $m_{\rm proton} n(z_{0}) R^{2}(z_{0}) a_{\rm s}(z_{0})$, where $m_{\rm proton}=1.67 \times 10^{-27}$~kg is the mass of the proton, $R(z_{0})$  the distance of the base of the wind from the centre of the planet, and $a_{\rm s}(z_{0})$ the sound speed at level $z_{0}$. For convenience,  Eq.~(64) of \citet{Adams11} is reproduced here as:
\begin{equation}
\dot{m} = 4.8 \, b^{3} \zeta^{1/3} \exp(-b),
\label{adams64}
\end{equation}
where $b \equiv G M_{\rm p}/[a_{\rm s}^{2} R(z_{0})]$ is a non-dimensional measure of the depth of the planet gravitational well and the equation is valid for $\zeta \la 0.01$.  { Eq.~(\ref{adams64}) assumes that the gravitational field of the planet effectively limits the evaporation rate as it is the case for a Jovian mass body. However, when less massive planets are considered, the parameter $b$ becomes lower because the planet radius  changes only slightly, thus reducing the energy required to accelerate the flow and moving the sonic point closer to the planet. When $b=3$ the sonic point along the polar streamline reaches the surface of the planet \citep[cf. Eq.~50 of ][]{Adams11} and Eq.~(\ref{adams64}) is no longer valid. In this regime, the wind can escape freely along field lines with an initial speed  comparable with the sound speed \citep[cf. the similar case of the free photoevaporation of circumstellar discs in ][]{StorzerHollenbach99,Adamsetal04a}. In conclusion, we assume that Eq.~(\ref{adams64}) is valid for $b > 3$, while we assume that $\dot{m} = 6.45  \, \zeta^{1/3}$ (i.e., is constant at its  maximum value reached at $b=3$) when $b \leq 3$ to avoid a  discontinuity in the variation of the mass loss rate as a function of the parameter $b$. 
}

Our simple wind model is based on an approximate treatment of the energy balance through the flow. Therefore, it comes of no surprise that it satisfies energy conservation only in order of magnitude. Specifically, we find that in $\approx 20$~percent of the cases, the mass loss rate exceeds its energy-limited value: 
\begin{equation}
\dot{M}_{\rm max} = \frac{\pi R_{\rm p}^{3} F_{\rm tot}}{G M_{\rm p}}, 
\label{dotm-max}
\end{equation}
where $F_{\rm tot}=F_{\rm EUV} + F_{0}$ is the total energy input to the planetary atmosphere per unit surface and time as given by the sum of the EUV and non-thermal electron fluxes \citep[cf., e.g., Eq.~(14) of ][]{Lecavelierdesetangs07}. The discrepancy is always 
smaller than a factor of $3-4$, so we can trust our model in order of magnitude and apply it to compare cases with and without the contribution of non-thermal electron energization.

\subsection{Evaporation rate in the case of an interconnecting loop}
\label{strong-interaction}

In Sect.~\ref{star-planet_inter}, the power released in  a stressed magnetic loop interconnecting the planet with the star was estimated and values about $2-3$ orders of magnitude larger than in the case of reconnection at the boundary of the planetary magnetosphere were obtained. With such a strong energization, the response of the planetary atmosphere cannot be predicted by  our simple isothermal wind model in view of the limitations discussed in Sect.~\ref{wind-model}. Moreover, since the release of magnetic energy is probably distributed all along the loop with the highest powers where the field is stronger (cf. Sect.~\ref{star-planet_inter}), a prediction of the electron energy spectrum is rather complicated. On the other hand,  a simple energy-limited model seems to be more appropriate to estimate the evaporation rate  for a preliminary investigation of the effects on planetary atmospheres. Thus we apply Eq.~(\ref{energy-limited-evap}) to estimate the mass loss rate in the case of a stressed interconnecting loop. In this case, the energy flux is the sum of the EUV flux and of the (much greater) energy flux coming from magnetic dissipation that falls on the fraction $f_{\rm AP}$ of the planetary surface as given by Eq.~(\ref{adamsfap}).

\section{Results}
\label{results}

\begin{figure}
\centering{
\includegraphics[width=8cm,height=10cm,angle=90]{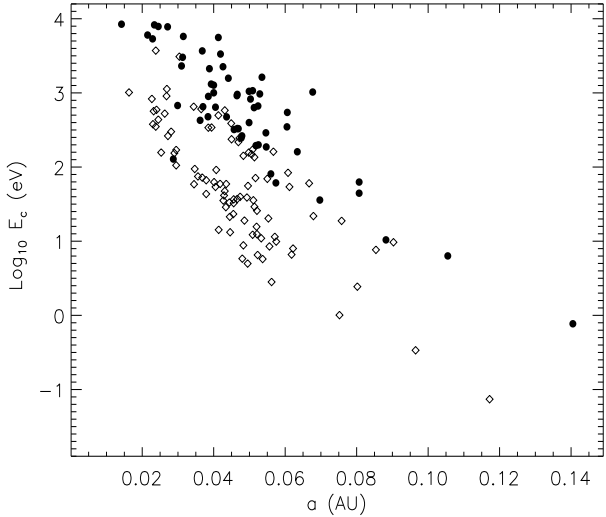}} 
\caption{The cutoff energy of non-thermal electron spectrum vs.  orbital semimajor axis for our sample of transiting planets for a spectrum with an index $\delta = 6$ and a stellar field decaying with distance with $s=3$ (cf. Eq.~\ref{stellar-field-var}). Solid dots refer to a stellar field $B_{0} \geq 10$~G, while 
open diamonds indicate $B_{0} < 10$~G (cf. Sect.~\ref{stellar-fields}). }
\label{electron_sp}
\end{figure}

In Fig.~\ref{electron_sp} we plot the cutoff energy $E_{\rm c}$ of the spectrum of non-thermal electrons accelerated at the boundary of the planetary magnetosphere. The cutoff energy is computed with the model in Sect.~\ref{electron-acceleration} assuming a stellar coronal base density of $5 \times 10^{14}$~m$^{-3}$, a coronal temperature $T_{\rm c}=10^{6}$~K \citep[cf., e.g., ][]{Poppenhaegeretal10}, and a spectral index $\delta=6$; the stellar field decays with distance as that of a dipole ($s=3$ in Eq.~\ref{stellar-field-var}). The accumulation of the plotted points along discrete sequences is due to the discrete values adopted for the stellar magnetic field as indicated by  different symbols (cf. Sect.~\ref{stellar-fields}). The dependence of $E_{\rm c}$ on the mass of the planet is quite weak and is not plotted. 

{ Close-in planets are characterized by values of $E_{\rm c}$ that  reach up to $\sim 10$~keV for $s=3$. The maximum cutoff energy reached in the case of a radial field dependence with $s=2$ in Eq.~(\ref{stellar-field-var}) is  $ \sim 120$~keV. This allows the penetration of  electrons  into the planetary atmosphere down to column densities ranging from  $10^{23}$ to $10^{25}$~m$^{-2}$ in the case of the most energetic spectra for $s=3$ and $s=2$, respectively. For $s=3$, the penetration depth is comparable with the level where most of the EUV photons are absorbed corresponding to a column density of $\sim 10^{22}-10^{23}$~m$^{-2}$. Only for the closest planets in the case with $s=2$, $E_{\rm c} \sim 100$~keV and the layer where most of the electron energy is deposited is remarkably deeper than that where EUV photons release most of their energy. In this regime, the parameter $\alpha_{\rm e} \equiv E_{\rm s}/E_{\rm c} \ll 1$ in the layer where the EUV photons are absorbed and the planetary wind is launched, thus allowing us to apply Eq.~(\ref{q-limit}) that gives an input energy rate approximately proportional to the number density $n(z)$. }

\begin{figure}
\centering{
\includegraphics[width=8cm,height=13cm,angle=0]{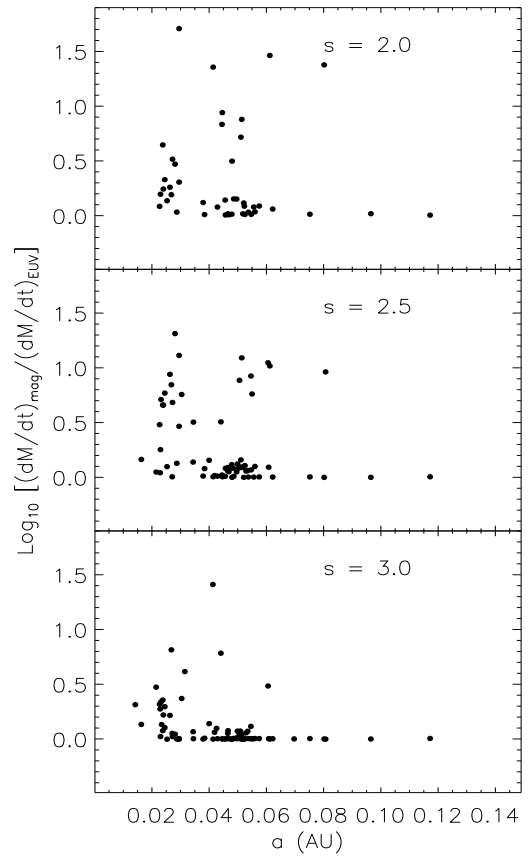}} 
\caption{Upper panel: Ratio of the mass loss rate induced by non-thermal electrons and EUV photons to that induced by EUV photons only  vs. the semimajor axis of the orbit for our sample of transiting planets (see the text). The parameter $s = 2.0$ specifies the radial dependence of the stellar magnetic field according to Eq.~(\ref{stellar-field-var}). Middle panel: the same as the upper panel, but for 
$s=2.5$. Lower panel: the same as the upper panel, but for $s=3.0$.  }
\label{ratiovsa}
\end{figure}

The  ratio of the mass loss rate produced by non-thermal electrons and  EUV flux to that induced by EUV flux only is plotted in Fig.~\ref{ratiovsa}
vs. the orbital semimajor axis for different values of the parameter $s$ determining the radial dependence of the stellar field in Eq.~(\ref{stellar-field-var}). 
We plot only transiting planets with  semimajor axis $a < 0.15$~AU,  mass $M _{\rm p}> 0.05$~M$_{\rm J}$,  with M$_{\rm J}$ being the mass of Jupiter, and $\zeta \leq 0.01$.  In the cases in which $\dot{M}_{\rm p}$ exceeds the energy-limited value  given by Eq.~(\ref{dotm-max}), the value of the mass loss rate has been fixed at that value (cf. Sect.~\ref{wind-model}).

Fig.~\ref{ratiovsa} can be compared with Fig.~\ref{pmagvspeuv} that showed the ratio of the corresponding input powers. Taking into account the limitations of our simplified model for planetary evaporation, we see that in general the ratio of the mass loss rates is comparable with or is a few times larger than that of the corresponding input powers. Magnetically-induced effects are larger for close-in planets and for smaller values of $s$, as expected on the basis of Fig.~\ref{pmagvspeuv}, and become small for  $ a \ga 0.1$~AU.  The mean increase of the mass loss rate is by a factor of $\sim 2.2$ for $s=2$,  by $\sim 2.1$, for $s=2.5$, and by $\sim 1.3$ for $s=3$, with a maximum relative increase by a factor of $30-50$ in the most extreme cases corresponding to closer-in  planets with mass  $\ga 1.5$~M$_{\rm J}$ (see below). 

\begin{figure}
\centering{
\includegraphics[width=8cm,height=13cm,angle=0]{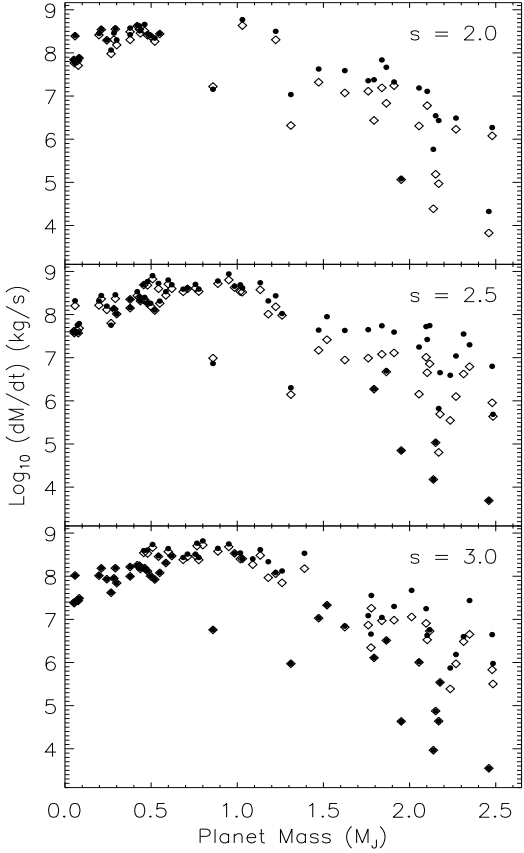}} 
\caption{Upper panel: The mass loss rate for a sample of transiting planets vs. their mass (see the text). Open diamonds indicate the loss rate induced by  EUV flux only, while filled dots indicate the case when the contribution by non-thermal electrons accelerated at the boundary of the planetary magnetosphere is added. The parameter $s = 2.0$ specifies the radial dependence of the stellar magnetic field according to Eq.~(\ref{stellar-field-var}). Middle panel: the same as the upper panel, but for 
$s=2.5$. Lower panel: the same as the upper panel, but for $s=3.0$.  }
\label{masslossrate}
\end{figure}

We plot in Fig.~\ref{masslossrate} the mass loss rate $\dot{M}_{\rm p}$ vs. the planet mass $M_{\rm p}$ with EUV energization only (open diamonds) and with both EUV and non-thermal electron energization (solid dots) for different values of $s$, respectively.
The overall dependence of $\dot{M}_{\rm p}$ on $M_{\rm p}$ is a consequence of the proportionality of $\dot{m}$ to $b^{3} \exp(-b)$ in Eq.~(\ref{adams64}) and the small variation in planet radius and sound speed at the base of the wind that makes $b$ approximately proportional to $M_{\rm p}$. A similar dependence is apparent in Fig.~9 of \citet{Adams11}, although his mass loss rates are more than one order of magnitude smaller than ours because of the systematically smaller EUV fluxes adopted by him and the assumption that only $\approx 0.5$ of the EUV flux goes into wind acceleration. Our values seem to be in better agreement with the observations of \object{HD~209458b} and \object{HD~189733b} for which $\dot{M}_{\rm p} \approx 10^{7}$ and $\approx 10^{8}$~kg~s$^{-1}$ have been reported, respectively (see Sect.~\ref{intro}). 

\begin{figure}
\centering{
\includegraphics[width=5.5cm,height=8cm,angle=90]{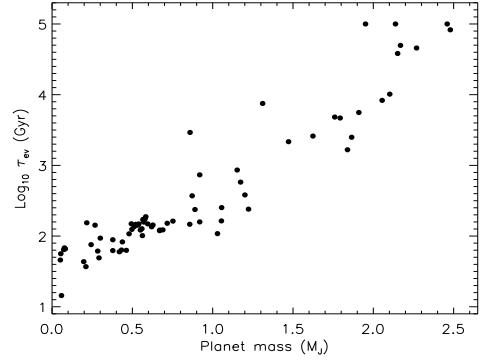}} 
\caption{Lifetime for the considered sample of transiting planets vs. their mass. The lifetime is estimated considering the maximum mass loss rate, i.e.,  by including the contribution of  non-thermal electrons accelerated by reconnection at the boundary of the planetary magnetosphere.    }
\label{lifetime}
\end{figure}

In none of the considered cases the magnetic increase of the mass loss  rate has a dramatic effect on the lifetime of the planets as estimated by $\tau_{\rm ev}= M_{\rm p}/\dot{M}_{\rm p}$. It is  plotted in Fig.~\ref{lifetime} vs. the planet mass. Assuming a constant mass loss rate, the cumulative effect after $\sim 5$~Gyr amounts to a total mass loss of at most $\approx 10$~percent. 

\begin{figure}
\centering{
\includegraphics[width=6cm,height=10cm,angle=90]{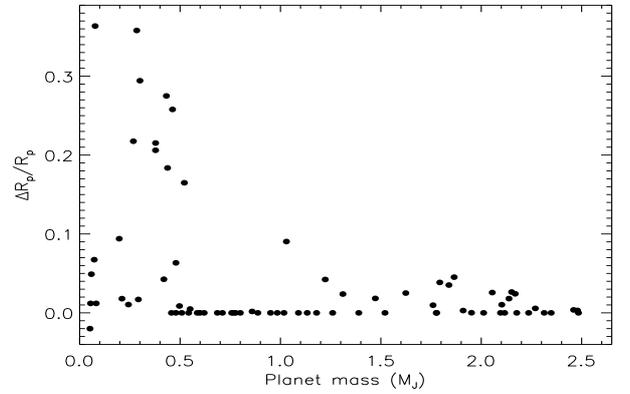}}  
\caption{Relative variation of the radius corresponding to optical depth $\tau =1$ in the centre of the Lyman-$\alpha$ line  for normal incidence across the planetary atmosphere vs. the planet mass. The plotted case corresponds to the  slowest decay of the stellar field with distance ($s=2.0$) that provides an upper limit to the predicted variation. }
\label{rad-variation}
\end{figure}

The non-thermal electron energy input can significantly lift the level corresponding to unity optical depth at UV wavelengths. This affects the  depth of the transit as  observed in UV lines. Although  a detailed comparison with the observations is outside the scope of the present investigation, we plot the relative variation of the radius corresponding to $\tau=1$ at the Lyman-$\alpha$ wavelength vs. the planet mass in Fig.~\ref{rad-variation}. For the systems with small magnetic effects, the variation is negligible, but for planets subject to a significant additional heating and having mass $M_{\rm p} \la 0.5$~M$_{\rm J}$, the increase can reach up to $30-40$~percent, implying a significantly deeper transit. Therefore,  the additional effects induced by star-planet magnetic interaction can be important for  modelling  transit observations in the UV.  

All the above estimates on the relative importance of non-thermal electron vs. EUV energy fluxes are conservative because we have considered an efficiency of $\sim 30$ percent in the conversion of magnetic energy into accelerated electrons with only half of the electron flux impinging upon the planet (cf. Eq.~\ref{magnetic-flux}). On the other hand, we do not apply any dilution factor to the EUV flux, although the acceleration of the wind happens only in the high-latitude regions of the planet. Depending on the parameter $\zeta$ in 
Eq.~(\ref{adamsfap}), the effective EUV flux can be reduced by a factor of $2-5$. 

\begin{figure}
\centering{
\includegraphics[width=8cm,height=8cm,angle=0]{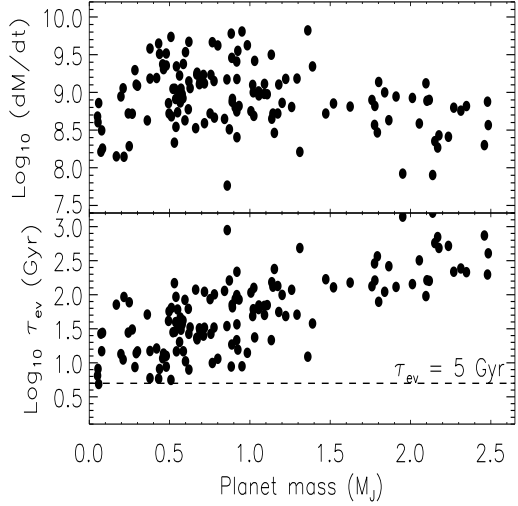}} 
\caption{Upper panel: mass loss rate averaged over time (in kg~s$^{-1}$) vs.  planet mass in the case of  the dissipation of magnetic energy in a loop  interconnecting the star with the planet. Lower panel: the estimated lifetime of evaporating planets  vs. their mass. The horizontal dashed line marks an evaporation lifetime of 5~Gyr corresponding to the mean estimated age of solar-like main-sequence stars. }
\label{strong_diss}
\end{figure}

Finally, we consider the possibility of a strong atmospheric heating in those systems in which there is a magnetic loop interconnecting the star with the planet. With the energy-limited approach of Sect.~\ref{strong-interaction}, the average mass loss rate and the lifetime of the planets vs. mass are shown in Fig.~\ref{strong_diss} that has been computed assuming a constant planetary field $B_{\rm p}=10$~G and a constant stellar field $B_{0}=2$~G with a radial dependence  with $s=2.5$.  The time-averaged mass loss rate is about one order of magnitude higher than in the case of  magnetic reconnection localized at the boundary of the planetary magnetosphere. The instantaneous mass loss rate can be one order of magnitude greater than the average plotted in Fig.~\ref{strong_diss}, computed by assuming that an interconnecting loop is present only $\sim 10$~percent of the time. With the adopted parameters, the lifetime of the transiting planets in our sample is always longer than 5~Gyr, indicating that a more frequent or stronger interaction is unlikely. 

\section{Conclusions}
\label{discussion}

We have investigated  star-planet magnetic interaction in close-in planets and estimated its impact on the evaporation of their atmospheres. We have considered the energy provided by  reconnection between planetary and stellar coronal fields, finding that it is comparable with or larger than the energy input due to stellar EUV radiation for close-in ($a \la 0.10$~AU) planets. A simple model to derive the energy spectrum of the electrons accelerated at the boundary of the planetary magnetosphere is introduced and the dissipation of their energy into the planet atmosphere is treated in detail. In most  close-in planets, accelerated electrons can reach atmospheric layers down to { column densities of $10^{23}-10^{25}$~m$^{-2}$, i.e., comparable with or significantly deeper than the layers where EUV photons are absorbed}, corresponding to column densities of $10^{22}-10^{23}$~m$^{-2}$. The heating by accelerated electrons is not uniformly distributed over the planetary surface, but is localized close to the magnetic poles at the footpoints of  magnetic fields lines, i.e., over an area of $\approx 10-20$ percent of the whole surface \citep{Adams11}.

The theory introduced above can be applied to model  evaporation or chemical reactions in the outer layers of a planetary atmosphere. Here, we limit ourselves to a simple application to the isothermal transonic wind model of \citet{Adams11} and \citet{Trammelletal11}. { The magnetic contribution is found to be important in close-by ($a \la 0.10$~AU), massive ($M_{\rm p} \ga 1.5$~M$_{\rm J}$) planets for which the evaporation rate can be increased up to a factor of $30-50$ in extreme cases. This is due to the additional heating of the atmosphere by the accelerated electrons that increases the temperature and the sound speed thus reducing the value of the parameter $b$ appearing in Eq.~(\ref{adams64}). In turn, the exponential dependence   on $b$ produces  a remarkable increase  of the mass loss rate  with the additional heating.  Moreover, this additional  atmospheric heating  produces  an increase of the apparent radius of the planet at EUV wavelengths that is relevant, e.g.,  to modelling the transit profile in the Lyman-$\alpha$ line.}

In addition to the energy released by magnetic reconnection at the boundary of the planetary magnetosphere, we consider   energy dissipation  in a loop interconnecting the stellar and planetary fields.  Such loops can be invoked to account for the hot chromospheric spots observed in some seasons in a few systems hosting hot Jupiters \citep{Shkolniketal05,Shkolniketal08}. The average mass loss rates predicted in that case reach up to $\approx (0.3 -1.0) \times 10^{10}$~kg~s$^{-1}$ and can have a remarkable impact on the lifetimes of close-in planets, possibly explaining the observed increase of the mean planet mass with the inverse of the semimajor axis in systems with close-in planets.  
 
In the present investigation, we have considered stars in the middle of their main-sequence evolution. Both surface magnetic fields and  EUV fluxes are larger in young stars, but the ratio of their  contributions to the heating of a planetary atmosphere is only weakly dependent on stellar age. Specifically, \citet{SanzForcadaetal11} propose that $F_{\rm EUV}$ scales as $\tau^{-1.24}$, where $\tau$ is the stellar age, while the magnetic field strength can be assumed to scale as $B_{0} \propto \tau^{-0.5}$ in order to account for  Skumanich's law of stellar rotation evolution \citep[cf., e.g.,  ][]{Kawaler88}. Since $P_{\rm mag} \propto B_{0}^{2}$, if $B_{\rm p}/B_{0}$ is assumed to be constant, the ratio $P_{\rm mag}/P_{\rm EUV} \propto \tau^{0.24}$, only weakly dependent on stellar age. However,  in pre-main-sequence  stars the EUV flux tend to saturate, while surface fields  $B_{0} \sim 10^{3}$ G are observed \citep[e.g., ][]{YangJohnsKrull11}.  In this phase, magnetic heating  dominates and leads to a remarkable evaporation of close-in planets, although  limited to a few tens of Myr. Therefore, a detailed investigation is needed to predict the impact on the present mass distribution of close-in planets.  
 
 Another characteristic of stellar magnetic fields is their  variability on widely different timescales that can account for the remarkable variation in mass loss rate observed in, e.g., HD~189733b \citep{Lecavelierdesetangsetal12}. On the other hand, on the basis of  the observed X-ray flux variability, the relative amplitude of the  EUV flux variability  is predicted to be smaller in stars more active than  the Sun \citep[cf., e.g., Sect.~6.1.3 of ][]{Claireetal12}, thus making it difficult to account for a large variability in the mass loss rate if evaporation is powered  solely by the EUV flux. 

\begin{acknowledgements}
The author wishes to thank an anonymous Referee for several valuable comments that allowed him to improve this work. He is grateful to Drs. A.~S.~Bonomo, A.~Collier~Cameron, S.~Desidera, R.~Gratton, I.~Pagano, E.~Shkolnik, and A. Vidotto for interesting discussions on star-planet interaction and planetary evaporation in systems hosting hot Jupiters. The use of solar spectra obtained from the Virtual Planetary Laboratory web site (http://depts.washington.edu/naivpl/content/models/solarflux/) is also gratefully acknowledged. 
\end{acknowledgements}

\appendix
\section{Treatment of relativistic electrons}
\label{relativity}
When the cutoff energy $E_{\rm c} \ga 0.2\, m_{\rm e} c^{2}$, the energy loss of the high-energy electrons during their motion through the plasma must be treated in a relativistic way. For energies up to 10~MeV,  losses by bremsstrahlung, synchrotron radiation, and Compton scattering are at least two orders of magnitude smaller than the collisional loss considered here \citep[cf., e.g., Fig.~1 in ][]{Haug04,McTiernanPetrosian90}. In the case of a fully ionized hydrogen plasma, the relativistic energy loss rate per electron is \citep[e.g. Eq.~(10) in ][]{Haug04}:
\begin{equation}
\frac{dE}{dt} = - 4 \pi r_{0}^{2} m_{\rm e} c^{3}\Lambda_{\rm ee}n  \beta^{-1},
\label{rel_loss}
\end{equation}
where $E$ is the kinetic energy of the particle, $r_{0} = e^{2}/m_{\rm e} c^{2}$ the classic electron radius, with $m_{\rm e}$ the electron mass and $c$ the speed of light, $\Lambda_{\rm ee}$ the Coulomb logarithm for electron-electron collisions, $n$ the total number density of protons, and $\beta \equiv v/c$ with $v$ the electron speed, not to be confused with the plasma $\beta$ used before. The kinetic energy is expressed in units of the rest energy as: $E \equiv \epsilon\, m_{\rm e} c^{2}$, with:
\begin{equation}
\epsilon = \frac{1}{\sqrt{1 - \beta^{2}}} -1. 
\end{equation}
Considering that $K^{\prime} \equiv 2 \pi e^{4} \Lambda_{\rm ee} $ and $dz = c \beta dt$, Eq.~(\ref{rel_loss}) can be recast in the form: 
\begin{equation}~
\frac{dE}{dz} = -2\frac{K^{\prime}}{E} n \frac{(1+\epsilon)^{2}}{\epsilon +2}, 
\label{energy-loss-rel}
\end{equation}
that represents the relativistic generalization of Eq.~(\ref{energy_loss}) and reduces to it when $\epsilon \ll 1$. On the other hand, for highly relativistic particles $\epsilon \gg 1$ and $dE/dz = -2(K^{\prime}/m_{\rm e} c^{2}) n$ becomes independent of the energy. Note that our simple treatment of  partial ionization by including a factor $(\lambda + x)$ in the r.h.s. of Eq.~(\ref{energy_loss}) is not rigorously valid in the relativistic regime \citep[cf., e.g., ][]{McTiernanPetrosian90}, but we shall extend its validity   given that  the introduced error in the energy loss rate does not exceed a factor of $4-5$  that can be absorbed into the other uncertainties of our model. 

To compute the energy loss per  proton $Q_{c}$, we need to specify the electron flux that is ${\cal F}_{0}(E_{0}) = c \beta {\cal S}(E_{0})$ in the relativistic regime. For electron energies $\epsilon \leq 0.2$ we shall continue to use Eq.~(\ref{dissipated-power}), while for greater energies we shall neglect the modification of the  spectrum produced by the loss of energy in the layers above the level $z$ because it is negligible at high energies.  Specifically, to stop an electron with an energy $E \simeq 0.2\, m_{\rm e}c^{2} \sim 100$~keV, we need a column density of $\sim 2.7 \times 10^{25}$~m$^{-2}$, that is much larger than the column density of $10^{22}-10^{23}$~m$^{-2}$ corresponding to the layer from which the evaporation flow is launched. Therefore, the adopted expression for $Q_{\rm c}$ in the relativistic regime taking into account Eq.~(\ref{energy-loss-rel}) is:
\begin{equation}
Q_{\rm c}(z)  =  K^{\prime \prime}  \left[ \int_{\alpha_{\rm e}}^{0.2 \alpha_{\rm r}} \frac{{\cal F}_{0}(w) dw }{\sqrt{w^2-\alpha_{\rm e}^{2}}} +  
2 c  \int_{0.2}^{\infty} \frac{1 +\epsilon}{\sqrt{\epsilon(\epsilon+2)}} {\cal S}(\epsilon) d \epsilon \right], 
\label{q-relat}
\end{equation}
where $K^{\prime \prime}$,  ${\cal S}$, $\alpha_{\rm e} \equiv E_{\rm s}(z)/E_{\rm c}$, and $w =  E_{0}/E_{\rm c}$ have already been introduced in the case of Eq.~(\ref{dissipated-power}), while $\alpha_{\rm r} \equiv  m_{\rm e} c^{2}/E_{\rm c}$. In the relativistic regime, the total electron energy flux used to compute $n_{\rm T}$ is given by the expression:
\begin{equation}
F_{0} = m_{\rm e} c^{3} \frac{\delta -1}{\delta} \left( \frac{n_{\rm T}}{\epsilon_{\rm c}} \right) {\cal I}, 
\end{equation}
where $\epsilon_{\rm c} \equiv E_{\rm c}/m_{\rm e} c^{2}$ and 
\begin{eqnarray}
{\cal I} & \equiv & \frac{1}{A} \int_{0}^{\infty} \epsilon \beta {\cal S}(\epsilon)  d\epsilon = \nonumber \\
           & = &  \int_{0}^{\epsilon_{\rm c}} \frac{\epsilon \sqrt{\epsilon (\epsilon +2)}}{1 + \epsilon} d \epsilon + 
            \int_{\epsilon_{\rm c}}^{\infty} \frac{\epsilon \sqrt{\epsilon (\epsilon +2)}}{1 + \epsilon} \left( \frac{\epsilon}{\epsilon_{\rm c}} \right)^{-\delta} d \epsilon. 
\label{I-integr}
\end{eqnarray}
All the integrals in Eqs.~(\ref{q-relat}) and (\ref{I-integr}) are evaluated numerically. 
Note that the expression for the normalization constant $A$ of the electron spectrum $\cal S$ derived from Eq.~(\ref{aconstant_norm}) is still valid in the relativistic regime, thus we can determine the cutoff energy $E_{\rm c}$ as in the classic regime. 

The asymptotic expression for the collisional ionization cross section is not modified in the relativistic regime \citep[cf., e.g., Eq.~(12) in ][]{Schlickeiseretal03}. Therefore, we continue to use our formulae by adopting the relativistic expression for the electron flux, i.e., $\tilde{\cal F} (E) = c \beta {\cal S}(E)$.

\end{document}